\documentclass[aps,pra,showpacs,notitlepage]{revtex4-1}

\usepackage{graphicx}

\begin{document}

\title{Characterization of measurement uncertainties using the correlations between local outcomes obtained from maximally entangled pairs}

\author{Shota Kino}
\author{Taiki Nii}
\author{Holger F. Hofmann}
\email{hofmann@hiroshima-u.ac.jp}
\affiliation{
Graduate School of Advanced Sciences of Matter, Hiroshima University,
Kagamiyama 1-3-1, Higashi Hiroshima 739-8530, Japan
}

\begin{abstract}
Joint measurements of non-commuting observables are characterized by unavoidable measurement uncertainties that can be described in terms of the error statistics for input states with well-defined values for the target observables. However, a complete characterization of measurement errors must include the correlations between the errors of the two observables. Here, we show that these correlations appear in the experimentally observable measurement statistics obtained by performing the joint measurement on maximally entangled pairs. For two-level systems, the results indicate that quantum theory requires imaginary correlations between the measurement errors of $\hat{X}$ and $\hat{Y}$ since these correlations are represented by the operator product $\hat{X}\hat{Y}=i \hat{Z}$ in the measurement operators. Our analysis thus reveals a directly observable consequence of non-commutativity in the statistics of quantum measurements. 
\end{abstract}


\maketitle
%%--Introduction

\vspace{-0.5cm}

\section{Introduction}

Recently, advances in experimental possibilities have renewed the interest in the physics of measurement uncertainties \cite{Mir09,Iin11,Erh12,Roz12,Tan13,Bae13,Rin14,Hof14a,Bae14}. This topic of research actually has a long history going all the way back to Heisenberg's justification of the quantum formalism by the definition of uncertainty limits for the simultaneous determination of position and momentum. Consequently, most of the recent work has focused on the quantitative evaluation of uncertainties using either the original evaluation by variances \cite{Oza03,Bra13,Wer14} or the more recent concept of entropic uncertainties \cite{Bro09,Weh10,Bus14}. Crucially, all of these approaches are based on the statistical limitations imposed by the mathematics of the quantum formalism, following Heisenberg's implicit suggestion that it is impossible to obtain any direct experimental evidence of the physics that causes the appearance of uncertainties in the first place. However, the study and characterization of entanglement shows that quantum correlations can exceed the local limits of uncertainties \cite{Hof03,Opp10,Ber14}. It is therefore possible to observe otherwise hidden details of the measurement error statistics by using the uncertainty free correlations of entangled states as a reference. In the present paper, we show how the correlations between the errors that characterize a joint measurement of two non-commuting observables can be obtained by analyzing the experimental results from maximally entangled inputs. Specifically, we show that the reconstruction of error statistics is independent of the model used for the measurement process or for the physics of the input state, resulting in an operational procedure for the reconstruction of error statistics. 
For two-level systems, the complete statistics of measurement errors is obtained. Significantly, the error statistics for quantum measurements consistent with the standard formalism of measurement theory exceed the bounds for real-valued measurement errors, effectively resulting in a derivation of complex-valued error probabilities from the experimentally observed data.  

The starting point of our discussion is a specific form of a joint measurement of the non-commuting observables $\hat{A}$ and $\hat{B}$, where the results of the joint measurement correspond to the $d^2$ combinations of eigenvalues observed in separate projective measurements of the two observables. For this kind of measurement, the error statistics of one of the two observables is given by the conditional probabilities that are observed experimentally for eigenstate inputs of that observable. Measurement errors can then be described in qualitative terms, based on the notion that the outcome is either correct or incorrect. The marginal error probabilities for either $\hat{A}$ or $\hat{B}$ are directly observed in measurements of the eigenstate inputs, while the correlations between errors need to be reconstructed from inputs with known correlations. By representing the measurement in terms of a joint error probability, we are essentially comparing the quantum statistics of measurements directly with the analogous classical statistics, similar to the way in which the Wigner function describes quantum statistics as an analog of classical phase space distributions. The problem we need to solve is that the correlations between non-commuting properties in a general quantum state are themselves unknown. For this purpose, we introduce the entangled states as an input that probes the correlations between measurement errors of non-commuting observables. Specifically, we make use of the fact that maximal entangled states provide precise correlations between the values of all physical properties in the two systems, so that an error free measurement results in a well-defined relation between the outcomes in system 1 and the outcomes in system 2 for both $\hat{A}$ and $\hat{B}$. It is then possible to deduce the error statistics using only the assumption that the joint probabilities for relations other than the ones observed in precise measurements of either $\hat{A}$ or $\hat{B}$ are zero.
 
We apply our analysis of measurement statistics to the orthogonal Bloch vector components $\hat{X}$ and $\hat{Y}$ of a two level system and find that the only missing element in the description of the measurement process is the correlation between the errors in $\hat{X}$ and $\hat{Y}$. To obtain this correlation experimentally, we then introduce a maximally entangled state of two systems. Since the correlations of $\hat{X}$ and $\hat{Y}$ between the two systems are known, the application of joint measurements to both systems should ideally produce the same correlations. From the actual results, we can judge whether the occurrence of an error in only one of the two measurements changed the correlation or not. Since the changes in correlations are obtained for both $\hat{X}$ and $\hat{Y}$ in a single measurement, the measurement outcomes reveal the correlation between the occurrence of errors in the two local measurements, and the missing element of the error statistics can be derived from the experimental results. 

We can use the standard formalism of quantum mechanics to predict the results of joint measurements on maximally entangled states. Interestingly, the predicted experimental results violate the limits that apply to any positive valued (and hence classical) selection of error probabilities. In fact, the quantum result indicates that the error correlation must be given by an imaginary part in the error probabilities. When these complex error statistics are applied to reconstruct the quasi-probabilities of an arbitrary input state, the result is the Kirkwood-Dirac distribution that is also observed in weak measurements \cite{McCoy32,Kir33,Dir45}. Since this distribution corresponds to the expectation value of an ordered product of the projection operators, the imaginary parts are a direct consequence of non-commutativity and essentially identify the expectation values of commutation relations with imaginary statistical correlations. The experimentally observed correlations between the measurement outcomes for the local measurements thus indicate that non-commutativity describes imaginary statistical correlations between real-valued observables. 

To properly appreciate the significance of the results, it is necessary to keep in mind that the physics of measurement can only be understood in terms of experimentally observable evidence. The essential insight of the following discussion is that entanglement provides us with the necessary tool for the analysis of correlations between non-commuting properties. If the claim that non-positive probabilities are a necessary consequence of the experimental evidence seems to be somewhat odd, it should be kept in mind that the observation or preparation of a joint reality for the non-commuting properties in question is impossible. The following discussion is therefore aimed at connecting the experimental evidence with the physical properties in a way that identifies the relevant correlations without requiring any unphysical assumptions about the reality of the system.

\section{Error statistics of joint measurements}
\label{sec:errors}

The established formalism of quantum mechanics represents physical properties by operators in Hilbert space. The precise measurement of a physical property $\hat{A}$ is represented by a projection onto an eigenstate $\mid a \rangle$, where the probability of an outcome $a$ is given by the product trace of the projection operator $\mid a \rangle \langle a \mid$ and the density matrix $\hat{\rho}=\mid \psi \rangle \langle \psi \mid$ that described the state $\mid \psi \rangle$ of the system before the measurement. Importantly, a joint assignment of outcomes $a$ and $b$ for two physical properties $\hat{A}$ and $\hat{B}$ is only possible if the two physical properties have shared eigenstates. It is therefore impossible to define a joint measurement of $a$ and $b$ without introducing some form of measurement uncertainty, such that the probability $P(a,b)$ of a joint outcome is not given by an intrinsic joint probability $\rho(a,b)$ of the initial state $\hat{\rho}$. Nevertheless, we can design a joint measurement that yields a complete set of joint outcomes $(a,b)$ if we accept statistical errors in the joint measurement. 

In standard quantum theory, such a joint measurement is described by a positive operator-valued measure, $\hat{\Pi}_{a,b}$, so that the joint probability of the experimental measurement outcome is given by 
\begin{equation}
P_{\mathrm{exp.}}(a,b)=Tr\{\hat{\Pi}_{a,b}\hat{\rho}\}.
\end{equation}
Mathematically, this is a bilinear relation between the measurement and the state defined in the Hilbert space of the system. However, it is the purpose of the measurement to evaluate the physical properties $\hat{A}$ and $\hat{B}$ in terms of their precise measurement outcomes $a$ and $b$ that would have been obtained from the original input state $\hat{\rho}$. For this purpose, we should express the input state in terms of the set of outcomes $(a,b)$ associated with precise error-free measurements. Here, it is interesting to observe that the quantum state $\hat{\rho}$ in a $d$-dimensional Hilbert space is described by $d^2$ independent parameters, corresponding to the number of possible combinations of measurement outcomes $a$ and $b$. Therefore, the only possible form of a bilinear relation of the measurement $\hat{\Pi}_{a,b}$ and the state $\hat{\rho}$ expressed in terms of the outcomes $a$ and $b$ is given by 
\begin{equation}
\label{eq:qprob}
P_{\mathrm{exp.}}(a,b)=\sum_{a^\prime,b^\prime} P(a,b|a^\prime,b^\prime)\rho(a^\prime,b^\prime),
\end{equation}
where the representation of the quantum state correspond to a joint probability $\rho(a,b)$ of the outcomes $a$ and $b$, and the description of the measurement process corresponds to a conditional probability relating the input combination $(a^\prime,b^\prime)$ to the measurement result $(a,b)$.
Importantly, Eq.(\ref{eq:qprob}) is not based on a hidden variable model and does not require the assumption that the combination of $a$ and $b$ in $\rho(a,b)$ represent a joint reality of $a$ and $b$ before the measurement is performed. The purpose of jointly assigning $a$ and $b$ to the input is merely to obtain a formulation of the joint measurement $\hat{\Pi}_{a,b}$ in terms of the target observables $\hat{A}$ and $\hat{B}$ that is equally valid for any input state, including eigenstates of either $\hat{A}$ or $\hat{B}$. Effectively, Eq.(\ref{eq:qprob}) should be understood as a general formulation of measurement that is analogous to classical statistics, just like a Wigner distribution is analogous to a classical phase space distribution. Since our goal is the evaluation of the error statistics $P(a,b|a^\prime,b^\prime)$, we will try to make only minimal assumptions about the form of the quasi-probability $\rho(a,b)$. It should be noted that this is possible because the measurement errors can be reconstructed from only a limited selection of input states. In the following, the states we consider for this purpose are eigenstates of the observables and maximally entangled states. In both cases, the form of the joint probabilities $\rho(a,b)$ and $\rho(a_1,b_1;a_2b_2)$ is completely determined by the experimental evidence and a set of reasonable constraints that would also be valid in classical probability theories. By using entangled states, it is possible to derive the non-classical correlations between errors using only the apparently classical correlations observed in separate measurements of either $\hat{A}$ or $\hat{B}$. 

Much like the definition of quasi-probabilities for quantum states, the formulation of $P(a,b|a^\prime,b^\prime)$ is motivated by the similarity between the description of classical measurement errors and the errors directly observed when the input value of either $\hat{A}$ or $\hat{B}$ is known. The procedure developed in the following allows a reconstruction of the quasi-probability $P(a,b|a^\prime,b^\prime)$ from experimental results that correspond to marginal probabilities of the complete error statistics. Similar to the reconstruction of a Wigner function from marginal distributions, it is natural that a reconstruction of the unobservable conditional probability $P(a,b|a^\prime,b^\prime)$ will result in non-positive quasi-probabilities. Nevertheless it should be kept in mind that these results are obtained from statistical assumptions about the relations between physical properties that appear to be completely consistent with all experimental observations. 

The argument is perhaps easiest to understand if we first look at the marginal distributions of $P(a,b|a^\prime,b^\prime)$ that are obtained by performing the measurement on a known eigenstate of $\hat{A}$ or $\hat{B}$.
In these cases, it is possible to identify a unique joint probability distribution of $a$ and $b$, since the eigenstates of $\hat{A}$ or $\hat{B}$ assign a probability of zero to any outcome other than the one specified by the state. The joint probabilities for the correct outcome are then given by the marginal probabilities for the eigenstates of the other property, as given by the squared inner products $|\langle b \mid a \rangle|^2$. For an eigenstate $\mid a \rangle$, the joint probabilities $\rho_a (a^\prime, b^\prime)$ are all zero for $a^\prime \neq a$ and correspond to the probabilities of $b^\prime$ in $\mid a \rangle$ otherwise,
\begin{equation}
\label{eq:a}
\rho_a(a^\prime,b^\prime)= \delta_{a,a'}|\langle b^\prime \mid a \rangle|^2. 
\end{equation}
Likewise, the eigenstates of $\hat{B}$ are described by joint probabilities of 
\begin{equation}
\label{eq:b}
\rho_b(a^\prime,b^\prime)= \delta_{b,b'} |\langle b \mid a^\prime \rangle|^2. 
\end{equation}
Importantly, we assume that there are no hidden non-classical probabilities for values of $a^\prime$($b^\prime$) other than the eigenvalue $a$($b$). Specifically, a quasi-probability $\rho_a(a^\prime,b^\prime)$ for an eigenstate $\mid a \rangle$ could also be formulated in such a way that the marginal for $a^\prime \neq a$ is zero because positive and negative quasi-probabilities cancel. Strictly speaking, the assumption that the value of $\hat{A}$ in an eigenstate $\mid a \rangle$ does not depend on the value of $\hat{B}$ is therefore one of the constraints that we choose to make a reconstruction of the total measurement statistics possible. However, it should be kept in mind that we are trying to find the closest possible analogy between quantum statistics and classical statistics, so the assumption of any joint probability other than the ones given in Eqs.(\ref{eq:a}) and (\ref{eq:b}) would seem to be rather artificial, given that classical statistics is sufficient to fully explain the experimental results in that specific case.

We can now characterize the conditional probabilities $P(a,b|a^\prime,b^\prime)$ experimentally by measuring the probabilities of the outcomes $(a,b)$ for input states of $\mid a^{\prime\prime} \rangle$ and $\mid b^{\prime\prime}\rangle$. The result is an intuitive description of measurement errors in terms of conditional probabilities that relate the correct input values to the actual output values. However, the fact that we have to select either eigenstates of $\hat{A}$ or eigenstates of $\hat{B}$ for the input means that we do not obtain information on the detailed correlations between the measurement errors $(a^{\prime\prime} \to a)$ and the measurement errors $(b^{\prime\prime} \to b)$. The experimental data only provides us with the input marginals of $P(a,b|a^\prime,b^\prime)$ given by 
\begin{eqnarray}
P_{\mathrm{exp.}}(a,b|a^{\prime\prime}) &=& \sum_{b^\prime} P(a,b|a^{\prime\prime},b^\prime) |\langle b^\prime \mid a^{\prime\prime} \rangle|^2,
\nonumber \\
P_{\mathrm{exp.}}(a,b|b^{\prime\prime}) &=& \sum_{a^\prime} P(a,b|a^\prime,b^{\prime\prime}) |\langle b^{\prime\prime} \mid a^\prime \rangle|^2.
\end{eqnarray}
Eigenstate inputs therefore permit only an incomplete characterization of the measurement statistics. To characterize the complete set of correlated errors described by the measurement operator $\hat{\Pi}_{a,b}$, it is necessary to use input states with well-defined correlations between the non-commuting properties $\hat{A}$ and $\hat{B}$. This is a non-trivial problem, since there is no clear consensus on the description of non-classical correlations by quasi-probabilities, and it is usually thought that such descriptions can only be obtained in the form of specific statistical models \cite{Spe07,Bar12}. Fortunately, this problem can be sidestepped by using maximally entangled states, and we will show in section \ref{sec:EPR} how correlations between the measurement errors can be identified in the experimental statistics obtained from correlated measurement of maximally entangled pairs. Before moving on to entanglement, however, it may be useful to take a closer look at the specific case of two-level systems, where it is sufficient to characterize the measurement outcome as either correct or incorrect.

\section{Evaluation of measurement errors in two-level systems}
\label{sec:flip}

The most simple example of a pair of non-commuting observables is give by the orthogonal spin components $\hat{X}$ and $\hat{Y}$ of a two-level system, where the eigenvalues are given by $\pm 1$. A joint measurement of $\hat{X}$ and $\hat{Y}$ therefore has four possible outcomes given by $(x,y)= (+1,+1), (+1,-1), (-1,+1), (-1,-1)$. As explained above, the joint measurement is described by 16 conditional probabilities $P(x,y|x^\prime,y^\prime)$ that relate the input values of $(x^\prime,y^\prime)$ to the output values $(x,y)$. However, the number of unknowns can be greatly reduced if we assume that the error probabilities are symmetric under exchanges of $+1$ and $-1$, so that the probability of obtaining a correct outcome does not depend on the actual spin value. The measurement statistics is then described by only four error probabilities defined by the relation between the input values and the output values. We can define these error probabilities as
\begin{eqnarray}
\label{eq:eta}
\eta(0,0) &=& P(x,y|x,y)
\nonumber \\
\eta(0,1) &=& P(x,-y|x,y)
\nonumber \\
\eta(1,0) &=& P(-x,y|x,y)
\nonumber \\
\eta(1,1) &=& P(-x,-y|x,y),
\end{eqnarray}
where the arguments of the error probabilities $\eta(r_x,r_y)$ define whether an error occurs ($r_i=1$) or not ($r_i=0$). Experimentally, it is always possible to satisfy this condition by randomly flipping the spin direction in the input, so that any differences in the errors for the four different input combinations of $x$ and $y$ average out in the overall measurement statistics. The removal of any experimental bias in $x$ or $y$ is therefore only a technical problem with a sufficiently simple solution.  

Following the procedure outlined in section \ref{sec:errors} we can now characterize the error probabilities $\eta(r_x,r_y)$ from experimental data obtained with eigenstate inputs. For the $\hat{X}$-eigenstates, the experimental results are
\begin{eqnarray}
P_{\mathrm{exp.}}(x,y|x) &=& \frac{1}{2} \left(\eta(0,0) + \eta(0,1)\right)
\nonumber \\
P_{\mathrm{exp.}}(-x,y|x) &=& \frac{1}{2} \left(\eta(1,0) + \eta(1,1)\right)
\end{eqnarray}
and these two results can be summarized by a single resolution parameter,
\begin{eqnarray}
\label{eq:V_x}
V_x &=& \sum_y \left( P_{\mathrm{exp.}}(x,y|x) - P_{\mathrm{exp.}}(-x,y|x) \right) 
\nonumber \\
&=& \eta(0,0) + \eta(0,1) - \eta(1,0) - \eta(1,1).
\end{eqnarray}
The resolution parameter $V_x$ represents the statistical contrast or visibility between the correct outcome of $x$ and the opposite outcome. $V_x=1$ denotes a precise measurement and $V_x=0$ describes a completely random measurement outcome, uncorrelated with the input. The results obtained from $\hat{Y}$-eigenstates can be summarized in the same manner, by 
\begin{eqnarray}
\label{eq:V_y}
V_y &=& \sum_x \left( P_{\mathrm{exp.}}(x,y|y) - P_{\mathrm{exp.}}(x,-y|y) \right) 
\nonumber \\
&=& \eta(0,0) - \eta(0,1) + \eta(1,0) - \eta(1,1).
\end{eqnarray}
The resolution parameters $V_x$ and $V_y$ can be evaluated directly by performing the measurement on eigenstate inputs and taking the difference between the probability of the correct measurement result and the probability of an error in the measurement of the known input observable. In addition, normalization requires that
\begin{equation}
\eta(0,0) + \eta(0,1) + \eta(1,0) + \eta(1,1) = 1.
\end{equation}
This leaves us with a single unknown parameter, which can be expressed by 
\begin{equation}
C = \eta(0,0) - \eta(0,1) - \eta(1,0) + \eta(1,1). 
\end{equation}
This expression describes the correlation between errors in $\hat{X}$ and errors in $\hat{Y}$ by distinguishing whether the errors occur jointly or separately. To evaluate this parameter, it is therefore necessary to know the correlations between $\hat{X}$ and $\hat{Y}$ in the input state, and this makes it necessary to consider the case of a maximally entangled input. 

\section{Correlations between entangled pairs} 
\label{sec:EPR}

Maximally entangled pairs are states with particularly strong correlation between two quantum systems. Specifically, a precise measurement of one system will project the remote system into an eigenstate corresponding to the measurement result obtained locally. If the physical properties of the two systems are properly aligned, the value of $\hat{A}$ in the remote system can be determined by an $\hat{A}$-measurement on the local system, and the value of $\hat{B}$ in the remote system is determined from the result of a $\hat{B}$-measurement on the local system. Maximally entangled pairs are therefore characterized by correlations in both $\hat{A}$ and $\hat{B}$, providing possible experimental evidence about the quantum correlations between the non-commuting observables $\hat{A}$ and $\hat{B}$. 

To obtain a joint probability for entangled pairs, we make use of the fact that the correlations observed in $\hat{A}$ and in $\hat{B}$ are completely consistent with classical statistics. As for the case of eigenstates given above, we assume that no non-positive probabilities should be introduced if the measurement statistics can also be explained in terms of conventional positive valued probabilities. In the case of maximally entangled states, this results in a well-defined joint probability for the measurement outcomes of $\hat{A}$ and $\hat{B}$. However, it should be noted that the values obtained for $\hat{A}$ and for $\hat{B}$ in both systems are not normally the same, so it may be important to consider the specific map between the values in system $1$ and the values in system $2$. In the following, we will represent this map by a tilde above the variable in system $1$, which indicates that the value represented by the expression is actually that observed in system $2$ as a consequence of the result in system $1$, so that $a_2=\tilde{a}_1$ and $b_2=\tilde{b}_1$. The joint probabilities of precise measurements in the two systems for the maximally entangled state $\mid E \rangle$ can then be given by 
\begin{eqnarray}
\label{eq:EC}
P(a_1,a_2) &=& |\langle a_1,a_2 \mid E \rangle|^2 = \frac{1}{d} \delta_{\tilde{a}_1,a_2}
\nonumber \\
P(b_1,b_2) &=& |\langle b_1,b_2 \mid E \rangle|^2 = \frac{1}{d} \delta_{\tilde{b}_1,b_2}.
\end{eqnarray}
It is also possible to determine the probabilities for a measurement of $\hat{A}$ in one system and a measurement of $\hat{B}$ in the other,
\begin{eqnarray}
\label{eq:Ep}
P(a_1,b_2) &=& |\langle a_1,b_2 \mid E \rangle|^2 = \frac{1}{d} |\langle \tilde{a}_1 \mid b_2 \rangle|^2
\nonumber \\
P(b_1,a_2) &=& |\langle b_1,a_2 \mid E \rangle|^2 = \frac{1}{d} |\langle a_2 \mid \tilde{b}_1 \rangle|^2.
\end{eqnarray}
Given the marginal probabilities above, it is possible to derive the complete joint probability of the maximally entangled state without any further assumptions. Specifically, Eq.(\ref{eq:EC}) indicates that all joint probabilities are zero unless $a_2=\tilde{a}_1$ and $b_2=\tilde{b}_1$, and Eq.(\ref{eq:Ep}) provides the specific probabilities of all non-zero contributions. Therefore, there is only one possible quasi-probability describing the maximally entangled state,
\begin{equation}
\rho_E(a_1,b_1;a_2,b_2) = \frac{1}{d} \delta_{\tilde{a}_1,a_2}\delta_{\tilde{b}_1,b_2}
|\langle a_2 \mid b_2 \rangle|^2.
\end{equation}
With this quasi-probability, it is possible to describe the correlated statistics of two independent measurements performed separately on the two systems. Here, we assume that it is possible to perform exactly the same measurement independently on system 1 and on system 2. Experimentally, this may be somewhat difficult, since the measurement setups may have different imperfections. Ideally, one would use the same setup twice using a time delay that is sufficiently long to reset the measurement system. If this is not possible, it would be important to ensure that both measurement setups are characterized by the same measurement statistics, perhaps even by randomly exchanging the roles of system 1 and system 2. Importantly, it is the goal of the measurement to characterize the joint measurement of $\hat{A}$ and $\hat{B}$ on a single system, so it is an experimental requirement of the present method of analysis that the measurement we wish to analyze can be applied independently to both systems. The experimental statistics of such an independent application of the measurement to system 1 and system 2 are described by the corresponding products of the two error probabilities, resulting in a quadratic function of the conditional probabilities $P(a,b|a^\prime,b^\prime)$ that describe the error statistics of the joint measurement of $\hat{A}$ and $\hat{B}$. Specifically,
\begin{equation}
\label{eq:corrstat}
P_{\mathrm{exp.}}(a_1,b_1;a_2,b_2|E) = \sum_{a^\prime,b^\prime}
P(a_1,b_1|a^\prime,b^\prime) P(a_2,b_2|\tilde{a^\prime}, \tilde{b^\prime})
\frac{1}{d} |\langle a^\prime \mid b^\prime \rangle|^2.
\end{equation}
In general, this is a very different sum from the one that determines the errors for eigenstate inputs. In particular, the sum runs over all $a^\prime$ and all $b^\prime$ of the input, since the input values of $\hat{A}$ and $\hat{B}$ are equally unknown. Nevertheless the correlations between system $1$ and system $2$ provide a clear statistical structure to the output, thus revealing something about the correlations between the measurement errors in $\hat{A}$ and in $\hat{B}$ in the form of correlations between the output values of $(a_1,b_1)$ and $(a_2,b_2)$.

\section{Error correlations for two-level systems}

We can now take a closer look at the case of singlet entanglement between a pair of two-level systems. In that case, all of the spin components have opposite values, so that $\tilde{x}_1=-x_1$ and $\tilde{y}_1 = - y_1$ and the joint probability of $x_i$ and $y_i$ can be written as
\begin{equation}
\rho_E(x_1,y_1;x_2,y_2)= \frac{1}{4} \delta_{-x_1,x_2} \delta_{-y_1,y_2}.
\end{equation} 
The experimental statistics for the correlated outcomes of joint measurements of system $1$ and system $2$ is given by 
\begin{equation}
\label{eq:2corr}
P_{\mathrm{exp.}}(x_1,y_1;x_2,y_2|E) = \frac{1}{4} \sum_{x^\prime,y^\prime}
P(x_1,y_1|x^\prime,y^\prime) P(x_2,y_2|-x^\prime, -y^\prime),
\end{equation}
where it is assumed that the error statistics of the two measurements are exactly identical. In an actual experiment, this could be accomplished by using the same measurement setup twice, with a sufficient time delay between the two measurements. Of course, some precautions should be taken to avoid cross-talk between the measurements, since this could induce artificial correlations. In setups where sufficient temporal separations between the measurements is not an option, care must be taken that the differences in the measurement procedures are completely random and uncorrelated, so that differences between the measurements merely add to the statistical errors described by $P(x_1,y_1|x^\prime,y^\prime)$. In this context, it may also be worth noting that experimental imperfections in the entanglement source can be overcome either by subtracting the statistical background noise associated with non-maximal correlations, or by simply attributing the errors in the entanglement source to the measurement process. In any case, it should not be too difficult to confirm the main results of the present paper experimentally even when the experimental setup does not achieve high levels of quantum coherence.

The sums in Eq.(\ref{eq:2corr}) can be further simplified by using the error probabilities $\eta(r_x,r_y)$ introduced in section \ref{sec:flip}. This means that the 16 experimental probabilities can also be summarized in terms of only four output patterns, depending on whether the observed correlations between $(x_1,y_1)$ and $(x_2,y_2)$ correspond to the original correlations or not:
\begin{eqnarray}
E(0,0) &=& P_{\mathrm{exp.}}(x,y;-x,-y)
\nonumber \\
E(0,1) &=& P_{\mathrm{exp.}}(x,y;-x,y)
\nonumber \\
E(1,0) &=& P_{\mathrm{exp.}}(x,y;x,-y)
\nonumber \\
E(1,1) &=& P_{\mathrm{exp.}}(x,y;x,y).
\end{eqnarray}
Each of the four values $E(r_x,r_y)$ can be determined experimentally by applying the joint measurement of $\hat{x}$ and $\hat{Y}$ to both systems of the entangled pair. According to Eq.(\ref{eq:2corr}), these experimental results are related to the error probabilities $\eta(r_x,r_y)$ by 
\begin{eqnarray}
\label{eq:Vx}
V_x^2 = \left(\eta(0,0)+\eta(0,1)-\eta(1,0)-\eta(1,1)\right)^2 &=& 4 \left(E(0,0)+E(0,1)-E(1,0)-E(1,1)\right)
\\ 
\label{eq:Vy}
V_y^2 = \left(\eta(0,0)-\eta(0,1)+\eta(1,0)-\eta(1,1)\right)^2 &=& 4 \left(E(0,0)-E(0,1)+E(1,0)-E(1,1)\right)
\\
\label{eq:Csquare}
C^2 = \left(\eta(0,0)+\eta(0,1)-\eta(1,0)-\eta(1,1)\right)^2 &=& 4 \left(E(0,0)+E(0,1)-E(1,0)-E(1,1)\right)
\end{eqnarray}
While Eq.(\ref{eq:Vx}) and Eq.(\ref{eq:Vy}) merely reproduce the resolutions obtained from the measurements of eigenstate inputs, Eq.(\ref{eq:Csquare}) provides direct experimental evidence of the correlation between measurement errors in $\hat{X}$ and $\hat{Y}$. 

\section{Quantum theory of joint measurements}

It is important to note that the results for the measurement resolutions $V_x$, $V_y$, and $C$ obtained from the experimental data given by the outcome probabilities $E(r_x,r_y)$ according to Eqs.(\ref{eq:Vx}-\ref{eq:Csquare}) are completely independent of the statistical model used to explain the measurement process. In particular, Eq.(\ref{eq:Csquare}) represents an operational definition of the correlation between measurement errors in $\hat{X}$ and $\hat{Y}$ that applies equally well to classical and to quantum models. However, quantum theory prevents an independent confirmation of the validity of Eq.(\ref{eq:Csquare}) by individual measurements of appropriately prepared inputs, since there are no joint eigenstates of $\hat{X}$ and $\hat{Y}$ that could be used to define the relation between the two non-commuting properties in the input. In quantum mechanics, Eq.(\ref{eq:Csquare}) thus represents the most fundamental operational definition of correlations between measurement errors in joint measurements of $\hat{X}$ and $\hat{Y}$.

Quantum theory actually makes very precise predictions about the measurement outcomes obtained from joint measurements of $\hat{X}$ and $\hat{Y}$. Since the measurement operators are defined by self-adjoint matrices in a two-dimensional Hilbert space, they can be expressed as linear combinations of the Pauli matrices $\hat{X}$, $\hat{Y}$, $\hat{Z}$ and the identity $\hat{I}$. The requirement of symmetry between results of $+1$ and results of $-1$ further limits the choice to equal coefficients with variable signs, so the only possible form of the positive operator-valued measure for a joint measurement of $\hat{X}$ and $\hat{Y}$ is
\begin{eqnarray}
\label{eq:POVM}
\hat{\Pi}_{+1,+1} &=& \frac{1}{4} \left(\hat{I} + V_x \hat{X} + V_y \hat{Y} + V_z \hat{Z}\right)
\nonumber \\
\hat{\Pi}_{+1,-1} &=& \frac{1}{4} \left(\hat{I} + V_x \hat{X} - V_y \hat{Y} - V_z \hat{Z}\right)
\nonumber \\
\hat{\Pi}_{-1,+1} &=& \frac{1}{4} \left(\hat{I} - V_x \hat{X} + V_y \hat{Y} - V_z \hat{Z}\right)
\nonumber \\
\hat{\Pi}_{-1,-1} &=& \frac{1}{4} \left(\hat{I} - V_x \hat{X} - V_y \hat{Y} + V_z \hat{Z}\right)
\end{eqnarray}
Here, $V_x$ and $V_y$ correspond directly to the experimentally observed resolutions for eigenstate inputs as discussed in sec. \ref{sec:flip} above. Since the correct outcomes should have probabilities greater than $1/2$, $V_x$ and $V_y$ should be positive numbers between zero and one. On the other hand, $V_z$ is not related to the experimental evidence discussed in sec. \ref{sec:flip}, so it represents an arbitrary parameter that can be either positive or negative. In addition, the quantum formalism requires that the operators are positive, so the sum of the squares of $V_i$ has a maximal value of one, corresponding to projections onto pure states. 

It is now possible to predict the experimental results that can be obtained from correlated measurements of maximally entangled pairs characterized by expectation values of $\langle \hat{X}_1 \hat{X}_2 \rangle =-1$, $\langle \hat{Y}_1 \hat{Y}_2 \rangle =-1$ and $\langle \hat{Z}_1 \hat{Z}_2 \rangle =-1$. The probabilities of the outcomes are given by the expectation values of the measurement operators,
\begin{eqnarray}
P_{\mathrm{exp.}}(x_1,y_1;x_2,y_2) &=& \langle \hat{\Pi}_{x_1,y_1} \otimes \hat{\Pi}_{x_2,y_2} \rangle
\nonumber \\ &=& \frac{1}{16}\left(1 - x_1 x_2 V_x^2 - y_1 y_2 V_y^2 - x_1 x_2 y_1 y_2 V_z^2   \right).
\end{eqnarray}
As expected, there are only four different results $E(r_x,r_y)$, where $r_x=0$ for $x_1=-x_2$ and $r_y=0$ for $y_1=-y_2$. Specifically, the four experimentally observable probabilities are given by 
\begin{eqnarray}
E(0,0) &=& \frac{1}{16}\left(1+V_x^2+V_y^2-V_z^2\right)
\nonumber \\
E(0,1) &=& \frac{1}{16}\left(1+V_x^2-V_y^2+V_z^2\right)
\nonumber \\
E(1,0) &=& \frac{1}{16}\left(1-V_x^2+V_y^2+V_z^2\right)
\nonumber \\
E(1,1) &=& \frac{1}{16}\left(1-V_x^2-V_y^2-V_z^2\right).
\end{eqnarray}
This result confirms the identification of $V_x$ and $V_y$ with the measurement resolutions of $\hat{X}$ and $\hat{Y}$ according to Eq.(\ref{eq:V_x}) and Eq.(\ref{eq:V_y}), respectively. Interestingly, the sensitivity to $\hat{Z}$ now appears as a correlations between the measurement errors in $\hat{X}$ and in $\hat{Y}$. However, this correlation is anomalous, since the sign of $C^2$ becomes negative:
\begin{eqnarray}
C^2 &=& \left(\eta(0,0)-\eta(0,1)-\eta(1,0)+\eta(1,1)\right)^2
\nonumber \\
 &=& 4 \left(E(0,0)-E(1,0)-E(0,1)+E(1,1)\right)
\nonumber \\
 &=& - V_z^2.
\end{eqnarray}
The experimentally observed correlations between measurement errors in $\hat{X}$ and $\hat{Y}$ therefore correspond to {\it imaginary} error probabilities defined by
\begin{equation}
C = \left(\eta(0,0)-\eta(0,1)-\eta(1,0)+\eta(1,1)\right) = \pm i V_z.
\end{equation}
This result clearly describes a qualitative difference between the predictions of quantum theory and the predictions of any possible classical model of measurement errors. Specifically, all classical statistical models would require that $E(0,0)+E(1,1)$ is greater than $E(0,1)+E(1,0)$. Indeed, it is obvious that a precise measurement of $\hat{X}$ and $\hat{Y}$ would have to result in the correct correlations, so that only $E(0,0)$ would obtain a non-zero value. The fact that all quantum measurements result in $E(0,1)+E(1,0) \geq E(0,0)+E(1,1)$ is therefore closely linked to the impossibility of uncertainty free joint measurements.

\section{Operator correlations and ideal measurements}

In order to understand why quantum theory can produce results that contradict classical expectations in a qualitative way, it is necessary to remember that quantum mechanics does not permit an uncertainty free joint measurement of the non-commuting properties $\hat{X}$ and $\hat{Y}$. If experimentally observable relations between non-commuting probabilities violate expectations associated with a hypothetical joint reality, the most likely conclusion is that there is no such reality. 

The present analysis allows us to identify the actual measurement errors in a quantum measurement, using a close analogy with classical error probabilities. The ideal classical measurement would be characterized by $\eta(0,0)=1$ and $V_x=V_y=C=1$. Interestingly, we can construct an operator that represents this ideal classical measurement by using $V_z=\pm i$ in Eq.(\ref{eq:POVM}). The resulting operators can then be factorized into a product of two projectors,
\begin{equation}
\label{eq:ideal}
\hat{\Pi}_{\mathrm{ideal}} = \frac{1}{4} \left(\hat{I}\pm\hat{X}\right)\left(\hat{I}\pm \hat{Y} \right).
\end{equation}
Quantum theory thus provides a well-defined theoretical form for uncertainty-free measurements, but this form is unphysical because it is non-hermitian and would therefore result in complex probabilities for the joint outcomes. Specifically, the correlations between $\hat{X}$ and $\hat{Y}$ are imaginary and satisfy the operator equation
\begin{equation}
\label{eq:imZ}
\langle \hat{X}\hat{Y} \rangle= i \langle \hat{Z} \rangle.
\end{equation}
Thus, our analysis suggests that (a) the correlation between $\hat{X}$ and $\hat{Y}$ is correctly represented by an ordered product of the operators, and hence has an imaginary value and (b) error free joint measurements are impossible because imaginary error probabilities are needed to convert the imaginary correlations into real-valued probabilities. 

Importantly, our analysis shows that the non-classical error correlations of joint measurements can be observed directly in experiments using entangled state inputs, without any prior assumptions about non-commutative observables. The correlations between the outcomes of joint measurements performed on entangled pairs thus show that the imaginary correlations represented by operator products do have experimentally observable consequences and should be taken seriously as valid descriptions of non-classical correlations. As Eq.(\ref{eq:ideal}) indicates, it is possible to invert the relation between experimental probabilities and the quasi-probability $\rho(x,y)$ to obtain the only possible definition of a joint probability for a two level system that is consistent with the experimentally observed statistics. This joint probability has the form of a Kirkwood-Dirac distribution \cite{McCoy32,Kir33,Dir45},
\begin{equation}
\label{eq:dirac}
\rho(x,y) = \mbox{Tr}\left(\hat{\Pi}_{\mathrm{ideal}} \hat{\rho} \right) = \langle x \mid y \rangle \langle y \mid \hat{\rho} \mid x \rangle.
\end{equation}
Significantly, this distribution has been derived by analyzing the errors of measurements for specific situations in which the form of $\rho(x,y)$ is not ambiguous and can therefore be formulated without having to chose between different quasi-probabilities. Our results therefore indicate that the only joint probability that is consistent with the experimentally observed correlations between measurement errors is the Kirkwood-Dirac distribution given by Eq.(\ref{eq:dirac}).

Consistency requires that the Kirkwood-Dirac distribution satisfies all of the assumptions we made when identifying the joint probability distributions of eigenstates and entangled states given in Eqs. (\ref{eq:a},\ref{eq:b},\ref{eq:corrstat},\ref{eq:2corr}). It is comparatively easy to confirm this for the case of the eigenstates. For entangled pairs, the Kirkwood-Dirac distribution is given by
\begin{eqnarray}
\rho_E(a_1,b_1;a_2,b_2) &=& \langle a_1,a_2 \mid b_1,b_2 \rangle \langle b_1,b_2 \mid E \rangle \langle E \mid  a_1,a_2 \rangle
\nonumber \\ &=& |\langle a_2 \mid b_2\rangle|^2 \frac{1}{d} \delta_{\tilde{b}_1,b_2} \delta_{\tilde{a}_1,a_2}.
\end{eqnarray}
Here, the delta functions originate from the same inner products of state vectors that also describe the experimentally observed statistics given in Eq.(\ref{eq:EC}). The Kirkwood-Dirac distribution therefore provides a description of entanglement that is particularly close to the classical explanation of the correlations between the properties $\hat{A}$ and $\hat{B}$ in the two systems. It may be worth noting that this result seems to be consistent with theoretical considerations about the conditions that define quasi-probabilities, where it was shown that the Kirkwood-Dirac distribution can be derived from a relatively small set of reasonable conditions \cite{Hof14c}. It may well be that the Kirkwood-Dirac distribution provides the best expression of the quantum correlations described by the Hilbert space formalism.

In the specific case of a two level system, the reconstruction of the Kirkwood-Dirac distribution from a joint measurement of $x$ and $y$ is particularly simple since it only involves the coefficients $V_i$. The reconstruction procedure essentially involves an inversion of these visibilities, where $C=i V_z$ is the imaginary visibility of the correlation. The inversion can then be given by 
\begin{eqnarray}
\label{eq:reconstruct}
\rho(x,y) &=& \frac{1}{4} \left(1 + \frac{1}{V_x} + \frac{1}{V_y} + \frac{1}{C}\right) P_{\mathrm{exp.}}(x,y)
\nonumber \\ && 
+ \frac{1}{4} \left(1 + \frac{1}{V_x} - \frac{1}{V_y} - \frac{1}{C}\right) P_{\mathrm{exp.}}(x,-y)
\nonumber \\ && 
+ \frac{1}{4} \left(1 - \frac{1}{V_x} + \frac{1}{V_y} - \frac{1}{C}\right) P_{\mathrm{exp.}}(-x,y)
\nonumber \\ && 
+ \frac{1}{4} \left(1 - \frac{1}{V_x} - \frac{1}{V_y} + \frac{1}{C}\right) P_{\mathrm{exp.}}(-x,-y).
\end{eqnarray}
The coefficients $V_x$, $V_y$ and $C$ are directly obtained from the experimental data of entangled states according to Eqs.(\ref{eq:Vx}), (\ref{eq:Vy}) and (\ref{eq:Csquare}), respectively. Since $C=i V_z$ is imaginary in two level quantum systems, Eq.(\ref{eq:reconstruct}) usually results in imaginary parts of the quasi-probability $\rho(x,y)$ when applied to the real and positive probabilities of the experimental measurement outcomes for $x$ and $y$. The experimental evidence obtained from entangled state inputs therefore indicates that the real and positive probabilities of joint measurement outcomes originate from a combination of complex joint probabilities with complex error probabilities, where the imaginary parts of the error probabilities are needed to compensate the imaginary correlations that are part of the non-classical statistics of the initial quantum states.

Although it is a widespread opinion that theoretical assumptions are necessary to formulate statistical theories of quantum mechanics, the present results suggest that these assumptions can be kept to a very reasonable minimum by considering the experimentally observable physics in detail. It is then possible to identify non-classical correlations directly in the experimental data, without the need for a specific statistical model. Interestingly, this is also consistent with recent progress in the analysis of quantum correlations using weak values, in particular the realization of direct measurements of the Kirkwood-Dirac distribution \cite{Lun11,Hof12a,Lun12}, which show that the Kirkwood-Dirac distribution is the only quasi-probability that correctly describes the non-classical correlations observed in weak measurements, and the observation that optimal cloning maps the unobservable correlations described by the Kirkwood-Dirac distribution onto observable correlations between the cloned outputs \cite{Hof12b,Bus15}. Given all of these results, it may well be possible that quantum mechanics can be explained completely by the experimental evidence obtained in properly evaluated measurements, and that interpretational difficulties are merely a result of misunderstandings caused by unnecessary abstractions in the mathematical formulation. In particular, it has already been shown by one of the present authors that the Hilbert space formalism can actually be derived completely from the experimental evidence obtained in weak measurements \cite{Hof14b}. The present results provide additional confirmation that these correlations are physical, and can be obtained from the experimental evidence without any arbitrary theoretical assumptions. 

\section{Conclusions}

Even though it is impossible to obtain uncertainty free values of non-commuting observables in a single joint measurement, it is possible to describe the statistical errors of joint measurements in terms of conditional probabilities relating the error-free results to the actual results. Since these statistical errors should not depend on the input state, they can be evaluated by using specific input states whose joint probabilities are known from the experimentally accessible statistics. These states are the eigenstates of the observables and maximally entangled states with perfect correlations in the two non-commuting properties. For two-level systems, the complete set of error probabilities can be obtained by comparing the correlations between the joint outcomes obtained for the two entangled systems with the known correlations of the initial entangled state. Since the analysis of the experimental results does not depend on any prior assumptions about the measurement statistics, it is possible to compare the predictions of quantum theory directly with the corresponding predictions of classical models of measurement errors.

Significantly, the experimentally observable correlations predicted by quantum theory are qualitatively different from the predictions of classical statistical models, since the correlations between the errors in $\hat{X}$ and $\hat{Y}$ correspond to imaginary error probabilities. This experimentally observable difference between quantum theory and classical statistics is a direct consequence of non-commutativity, since the imaginary probability can be traced to the operator product $\hat{X}\hat{Y} = i \hat{Z}$. Uncertainty limited joint measurements of non-commuting observables are therefore sensitive to the non-classical correlations described by the non-commutativity of the operators that represent physical properties in the quantum formalism. By performing the same type of joint measurement on two maximally entangled systems, it is possible to make these non-classical correlations visible in the form of joint probability distributions that cannot be explained in terms of positive-valued statistics. The correlations between the outcomes of joint measurements observed with entangled input states thus provides direct experimental evidence for the non-classical correlations associated with non-commutativity, indicating that non-commutativity corresponds to imaginary correlations between physical properties. By taking these experimentally observable features into account, it may be possible to develop a complete theory of quantum statistics without any arbitrary assumptions from speculative models of reality. 

\section*{Acknowledgements}
This work was supported by JSPS KAKENHI Grant Number 24540427.


\begin{thebibliography}{xyz00}

%%--which path weak measurement
\bibitem{Mir09}
R. Mir, J. S. Lundeen, M. W. Mitchell, A. M. Steinberg, J. L. Garretson, and H. M. Wiseman, New J. Phys. {\bf 9}, 287 (2007).

%%--sequential measurement
\bibitem{Iin11}
M. Iinuma, Y. Suzuki, G. Taguchi, Y. Kadoya, and H. F. Hofmann,
New J. Phys. {\bf 13}, 033041 (2011). 

%%--Ozawa related experiments
\bibitem{Erh12}
J. Erhart, S. Sponar, G. Sulyok, G. Badurek, M. Ozawa, and Y. Hasegawa, Nature Physics {\bf 8}, 185 (2012).

\bibitem{Roz12}
L. A. Rozema, A. Darabi, D. H. Mahler, A. Hayat, Y. Soudagar,
and A. M. Steinberg, Phys. Rev. Lett. {\bf 109}, 100404 (2012).

%%---complementarity
\bibitem{Tan13}
J.-S. Tang, Y.-L. Li, C.-F. Li, and G.-C. Guo, Phys. Rev. A {\bf 88}, 014103 (2013).

%%--Ozawa related experiments
\bibitem{Bae13}
S.-Y. Baek, F. Kaneda, M. Ozawa, and K. Edamatsu, Sci. Rep.
{\bf 3}, 2221 (2013).

\bibitem{Rin14}
M. Ringbauer, D. N. Biggerstaff, M. A. Broome, A. Fedrizzi,
C. Branciard, and A. G. White, Phys. Rev. Lett. {\bf 112}, 020401
(2014).

%%--quantum controlled measurement
Hof03
\bibitem{Hof14a}
H. F. Hofmann, New J. Phys. {\bf 16}, 063056 (2014).

%%--sequential measurements
\bibitem{Bae14}
K. Baek, T. Farrow, and W. Son, Phys. Rev. A {\bf 89}, 032108 (2014).

%%---Ozawa and alternatives

\bibitem{Oza03}
M. Ozawa, Phys. Rev. A {\bf 67}, 042105 (2003).

\bibitem{Bra13}
C. Branciard, Proc. Natl. Acad. Sci. U.S.A. {\bf 110}, 6742 (2013).

\bibitem{Wer14}
P. Busch, P. Lahti, and R. F. Werner, Rev. Mod. Phys. {\bf 86}, 1261 (2014).

%%--entropic uncertainty

\bibitem{Bro09}
T. Brougham, E. Andersson, and S. M. Barnett, Phys. Rev. A {\bf 80}, 042106 (2009).

\bibitem{Weh10}
S. Wehner and A. Winter, New J. Phys. {\bf 12}, 025009 (2010).

%%--information measure

\bibitem{Bus14}
F. Buscemi, M. J. W. Hall, M. Ozawa, and M. M. Wilde, Phys. Rev. Lett. {\bf 112}, 050401 (2014).

%%--measurement uncertainty and entanglement
\bibitem{Hof03}
H. F. Hofmann and S. Takeuchi, Phys. Rev. A {\bf 68}, 032103 (2003).

\bibitem{Opp10}
J. Oppenheim and S. Wehner, Science {\bf 19}, 1072 (2010).

\bibitem{Ber14}
M. Berta, P. J. Coles, and S. Wehner, Phys. Rev. A {\bf 90}, 062127 (2014).

%%--Kirkwood-Dirac

\bibitem{Kir33}
J. G. Kirkwood, Phys. Rev. {\bf 44}, 31 (1933).

\bibitem{McCoy32}
N. H. McCoy, Proc. Natl. Acad. Sci. U. S. A. {\bf 18}, 674 (1932).

\bibitem{Dir45}
P. A. M. Dirac, Rev. Mod. Phys. {\bf 17}, 195 (1945).  

%%%---statistical models
\bibitem{Spe07}
R. W. Spekkens, Phys. Rev. A {\bf 75}, 032110 (2007).

\bibitem{Bar12}
S. D. Bartlett, T. Rudolph, and R. W. Spekkens, Phys. Rev. A {\bf 86}, 012103 (2012).

%%%--Dirac dist
\bibitem{Hof14c}
H. F. Hofmann, Quantum Stud.: Math. Found. {\bf 1}, 39 (2014).

%%%---complex probabilities
\bibitem{Lun11}
J. S. Lundeen, B. Sutherland, A. Patel, C. Stewart ,and C. Bamber, Nature {\bf 474}, 188 (2011).

\bibitem{Hof12a}
H.F. Hofmann, New J. Phys. {\bf 14}, 043031 (2012). 

\bibitem{Lun12}
J. S. Lundeen and C. Bamber, Phys. Rev. Lett. {\bf 108}, 070402 (2012). 

\bibitem{Hof12b}
H. F. Hofmann, Phys. Rev. Lett. {\bf 109}, 020408 (2012).

\bibitem{Bus15}
F. Buscemi, M. Dall'Arno, M. Ozawa, and V. Vedral, Int. J. Quantum Inform. DOI: 10.1142/S0219749915600023 (2015).

\bibitem{Hof14b}
H. F. Hofmann, Phys. Rev. A {\bf 89}, 042115 (2014).

\end{thebibliography}
\end{document}